\documentstyle[12pt]{article}

\def\be{\begin{equation}}
\def\ee{\end{equation}}
\def\ba{\begin{array}{c}}
\def\ea{\end{array}}

\begin{document}

\titlepage
%\vspace*{4cm}

  \begin{center}
{\Large \bf
 Experiments in ${\cal PT}-$symmetric quantum mechanics
 }

\end{center}

\vspace{5mm}

  \begin{center}

Miloslav Znojil \vspace{3mm}

\'{U}stav jadern\'e fyziky AV \v{C}R, 250 68 \v{R}e\v{z}, Czech
Republic\\

e-mail: znojil@ujf.cas.cz

%\end{center}

\end{center}

\vspace{5mm}

\vspace{5mm}

\section*{Abstract}

Extended quantum mechanics using non-Hermitian (pseudo-Hermitian)
Hamiltonians $H = H^\ddagger$ is briefly reviewed.  A few related
{mathematical} {experiments} concerning supersymmetric
regularizations, solvable simulations and large-$N$ expansion
techniques are summarized. We suggest that they could initiate a
deeper study of nonlocalized structures (quasi-particles) and/or of
their unstable and many-particle generalizations. Using the
Klein-Gordon example for illustration we show how the ${\cal PT}$
symmetry of its Feshbach-Villars Hamiltonian $H^{(FV)}$ might clarify
{experimental} aspects of relativistic quantum mechanics.

\vspace{9mm}

\noindent PACS 03.65.Bz  03.65.Ca

 \vspace{9mm}
        % Classification Scheme
KEYWORDS: quantum mechanics; relativistic kinematics; non-Hermitian
observables
%  \begin{center}
% {\small \today, sh.tex file}
% \end{center}

%\newpage

%\title{ Experiments in ${\cal PT}-$symmetric quantum mechanics}
  % use lower case
%
          % Group together all authors with the same affiliation
          % & address and complete the commands `\author' and
          % `\address' subsequently for i, ii,..., vi.
          % Leave the remaining items untouched.
%\authori
%{Miloslav Znojil}
%    \addressi

%    {\'{U}stav jadern\'e fyziky AV \v{C}R, 250 68 \v{R}e\v{z},
%Czech Republic}

%
%Page headings:
%\headauthor{Miloslav Znojil}            % page heading on the even pages
%\headtitle{Experiments in ${\cal PT}-$symmetric quantum mechanics}
  % page heading on the odd pages
%\lastevenhead{M. Znojil:  Experiments in ${\cal PT}-$symm. quantum mechanics}
 % p. h. on the last page if even
%
%\pacs

% PACS {03.65.Bz  03.65.Ca}     % max. 2 codes of Physics and Astronomy

 % lowercase letters
%%%%%%%%%%%%%% FOR EDITORIAL USE ONLY!!! %%%%%%%%%%%%%%%
%\refnum{A}%\total{}\type{}
%\daterec{XXX}    %;\\ final version }
%\issuenumber{0}  \year{2001} \setcounter{page}{1}
%\firstpage{1}
%\lastpage{000}
%\makefirsttitle
%%%%%%%%%%%%%%%%%%%%%%%%%%%%%%%%%%%%%%%%%%%%%%%%%%%%%%%%
%\maketitle

%\begin

% \end{abstract}

\section{Introduction}

The formulation of ${\cal PT}-$symmetric quantum mechanics (PTSQM,
\cite{BB}) is not completed yet. The great portion of its intensive
recent development (sampled by contributions \cite{a1}-\cite{a21} in
this volume as well as by many further references cited therein) is
attracted by the related open questions in mathematics. New methods
are being developed using perturbation series and their resummations
\cite{a3,my}, sophisticated changes of variables \cite{a4,a8,a10},
semiclassical analysis \cite{a14,Alvarez}, pseudo-orthogonal random
matrix ensembles \cite{AJ} as well as efficient approximations
\cite{a5}, updated numerical techniques~\cite{a7,a15} and abstract
representation theory \cite{a16,a21}.

During all the similar speculations one must keep in mind that
physics is, undoubtedly, an experimental science. All its predictions
must be verified or falsified by measurements. Although the related
arguments are scattered over the literature, they support the natural
expectation that a ``being-the-real-physics" test will be passed by
at least some of the PTSQM-type theories. Still, the applicability
and applications of ${\cal PT}-$symmetric Hamiltonians in
phenomenology may seem, temporarily, less impressive, even though
several models already proved useful, reflecting, e.g., the
supersymmetry of a system \cite{a17,a18} or its finite-dimensional
Hilbert-space description \cite{a20}. Their not quite usual
properties inspired intensive old as well as new developments in
field theory \cite{a2,a11}, in nuclear physics \cite{a6,GK}, in
many-body theory \cite{a1,a5,a8} and, last but not least, in
condensed matter physics \cite{HN} and, recently, in
cosmology~\cite{a12} and magneto-hydrodynamics~\cite{uwe}.

The need of the necessary full balance between the steady progress in
mathematics and its fructification in experimental physics inspired
our present short review. In fact, there exist several {reasons for
changing an overall reluctance} of our colleagues in experimental
physics to weaken their reliance on the comfort provided by the
Hermitian phenomenological models. We mention just a few.

(1) In the general non-Hermitian context with the conservation of the
mere pseudonorm \cite{ich}, new types of observables should be
introduced and studied, having an indeterminate sign. One might, for
example, change the current habit of characterizing a point particle
by its mass (which is positive semidefinite) and try to measure,
e.g., the (conserved) charge in the fields or systems of charged
particles and antiparticles the mass of which could be neglected.

(2) One should note that in the PTSQM context, the laboratory of
solvable models is being enriched by many new completely as well as
partially solvable items; in particular, the superintegrable or
Calogerian models of many-body spectra become significantly modified
within the complexified, ${\cal PT}-$symmetric framework~\cite{a1}.
In all these exactly solvable models, a transition to their
non-Hermitian versions enables one to specify the charge itself anew,
as an operator of the so-called quasi-parity, which was introduced in
ref.~\cite{220} and given new emphasis in~\cite{BB2}.

(3) New horizons may be revealed when one tries to revisit
relativistic quantum theory. Recently, A. Mostafazadeh has noticed
that the free Klein-Gordon operator of coordinate acquires certain
interesting and quite subtle ``minimally nonlocal" properties
\cite{july}. This might add new interpretations to the famous EPR
paradox, one of the most influential ``Gedankenexperimente" which
made the nonlocality in quantum mechanics explictly exposed to a
broad public even in a non-relativistic setting.

(4) A similar paradox connected much more directly to the
relativistic kinematics re-emerges in Bethe-Salpeter equation which
offers a ``minimal" description of a (relativistic) two-body problem
in which only one time coordinate should remain observable. In such
an ``operator of time" context, a return to the nonrelativistic limit
leads to many non-particle concepts like, e.g., the quantum clock
\cite{time}. It is obvious that during the consistent physical
interpretation of all the similar systems one may significantly
deviate from the current versions and applications of the
correspondence principle.

% \noindent
In parallel to the new possibilities to describe physical reality one
must also solve numerous open questions in the formalism itself.
Giving a purely personal selection I would like to list, e.g., the
puzzles emerging within the framework of the ${\cal PT}-$symmetric
versions  of the popular large-$N$ expansion techniques
\cite{gerdt,low},  of the more-than-one-body exactly solvable models
\cite{low,angul} and of the use of the simplifying discretizations
\cite{angul,delta}.

The common purpose of all these (and similar) studies  should be a
verification of our understanding of the subtleties of the PTSQM
formalism of ref.~\cite{BB}.

\section{Less usual properties of pseudo-Hermitian Hamiltonians}

The difference between textbook Quantum Mechanics and the apparently
nonstandard formalism of PTSQM may be illustrated in a schematic
two-dimensional Hilbert space for which we may compare the real and
time-dependent Hermitian toy Hamiltonian
 \be
H_{(+)}(t)=\left(
\begin{array}{cc}
a&b(t)\\ b(t)&-a \ea \right ) \equiv H_{(+)}^\dagger(t)
 \label{twobytwoH}
 \ee
with its non-Hermitian, ${\cal PT}-$symmetric analogue
 \be
H_{(-)}(t)=\left(
\begin{array}{cc}
a&b(t)\\- b(t)&-a \ea \right )=\left(
\begin{array}{cc}
1&0\\0&-1\ea \right )\,H_{(-)}^\dagger\,\left(
\begin{array}{cc}
1&0\\0&-1\ea \right ) \equiv H_{(-)}^\ddagger(t).
 \label{twobytwoPT}
 \ee
The pertaining spectra are both given as roots of their secular
determinants,
 \be
 E_{(\pm)\,n}(t) =(-1)^n\, \sqrt{a^2 \pm b^2(t)}, \ \ \ \ n =
 1, 2.
 \ee
This indicates that the energies may be real not only in the common
Hermitian case, with interesting consequences in random matrix
theories~\cite{AJ} etc.

In the ${\cal PT}-$symmetric example $H_{(-)}(t)$ we have to
distinguish between three regimes, with $|a|> |b|$ (all energies are
real), with $|a| < |b|$ (all energies occur in complex conjugate
pairs) and with $|a| = |b|$ (``intermediate" domain). According to a
number of authors \cite{Dirac}-\cite{AM},
% \cite{Dirac,FV,textbook,Constantinescu,ich,BB2,AM}
this is a generic situation where one may simplify the discussion by
changing the scalar product in Hilbert space. In particular, in the
regime where all the energies remain real (i.e., at $|a|> |b|$ in our
illustrative $H_{(-)}(t)$), one can reinterpret the non-Hermitian
Hamiltonian as an operator which is Hermitian in the new metric.

Although the connection between the old and new metric may be
complicated in general, the puzzle is solved at least partially in
principle. In the new models with real spectra one need not modify
the foundations of quantum theory at all. A new territory is merely
found there for an innovative applications of the classical --
quantum correspondence.

In our toy model (\ref{twobytwoPT}) one can easily visualize, in
addition, the confluence of the two eigenvalues and, possibly, their
subsequent complexification at $t > t_0$. This may mean a {\em decay}
of our system. Within such an unorthodox quantum model based on the
weakened mathematical assumptions, the matrix form of the
non-Hermitian Hamiltonian would be permitted to contain the so-called
Jordan blocks. The occurrence of such a situation need not be
exceptional at all, finding its elementary illustration not only in
the spiked harmonic oscillator~\cite{220} but also in some much more
sophisticated models in field theory~\cite{Lee}.

In the next step of time evolution of any generic model, the
complex-conjugate pairs of the energies may emerge as generated by a
smooth change of the coupling strengths. Beyond the point of the
decay of the system, there may still exist an indirect access to the
spectra and wave functions in a non-causal domain. A few preliminary
steps in this promising direction have already been made in field
theory~\cite{KleefKy} as well as within supersymmetric quantum
mechanics~\cite{NPB}.

\section{The Klein-Gordon equation as a generic illustrative example}

% \subsection{Hamiltonian and its ${\cal PT}-$symmetry}

The use of non-Hermitian Hamiltonians $H$ seems incompatible with
Stone's theorem which relates the Hermiticity of $H$ to the unitarity
of the time evolution. A counterexample against such a current belief
is provided by Klein-Gordon equation for a spinless particle of mass
$m$ and charge $e$. In an external electromagnetic four-field denoted
by $A_\mu$ with index $\mu = 0, 1, 2, 3$ such a particle may be
assigned the Klein-Gordon operator~\cite{Greiner}
 \be
 H^{(KG)}=
 \left [\sum_\mu
 \left (
 p_\mu-\frac{e}{c}A_\mu
 \right )^2+m^2c^2
 \right ] \,\psi, \ \ \ \ \
 p_\mu=-i\hbar\frac{\partial}{\partial x_\mu}
 \label{lapta}
 \ee
in a way compatible with the requirements of relativistic covariance.

The core of the idea is that one does not need the standard form
$H^{(KG)}\,\psi=0$ of the Klein-Gordon equation itself but rather a
definition of the (instantaneous, i.e., in general, time-dependent)
Feshbach-Villars' \cite{FV} generator $H^{(FV)}$ of the time
evolution which happens to be different from $H^{(KG)}$ of course
[see, e.g., eq. (XII.13) in \cite{Constantinescu}]. The reason is
that the differential operator $H^{(KG)}$ is of the second order in
time.

The explicit definition of $H^{(FV)}$ may be copied from any textbook
[see, e.g., eq. (7.7) in \cite{Constantinescu}]. We only have to keep
in mind that this operator is given, by definition, just in a {\it
fixed} {inertial frame}. In this sense one has to reconsider the
meaning of the concept of observable quantities (like energy or
coordinate) and of the ``preparable" wave functions and of their
feasible ``filtration" under the relativistic kinematics.

An enormous advantage stems from the fact that {\em all} the
non-Hermitian operators $H^{(FV)}$ prove manifestly ${\cal
PT}-$symmetric after one defines the operator of ``parity" in terms
of the third Pauli matrix, ${\cal P} \equiv \sigma_3$. We recommend
the recently refreshed and updated presentation of the latter set of
problems by A. Mostafazadeh~\cite{july} who considers the free motion
in more detail, working with the simplified partitioned matrix
operator
 \be
 H^{(FV)}=-\frac{1}{2}(\sigma_3+i\sigma_2)\nabla^2+\sigma_3\,,
 \ee
where the speed of light $c$, Planck constant $\hbar$ and mass $m$
have all been put equal to one [cf. also eq. (2e)
in~\cite{Constantinescu}].

Many questions become re-opened in the light of the latter new
development. {\it Pars pro toto}, let us consider the relativistic
version of repeated measurements of a system in a state which has
been fixed (``prepared", projected on a ket $|\psi\rangle$) in a
distant past and which is measured again in a distant future
(``filtered", projected on the same ket $|\psi\rangle$). By the
postulates of quantum mechanics, the new measurement does not change
the wave function even if the ``past" and ``future" frames move with
respect to each other. In such an arrangement one arrives at a new
version of the old EPR paradox since one can hardly imagine a
consistent experimental setup in which the measurement would obey the
standard causality requirements.

\section{Summary and outlook}

Let us re-emphasize that the time development in PTSQM may move us to
a point $t=t_0$ at which our systems start living in an
``intermediate" regime. As long as the new metric in Hilbert space
becomes manifestly singular at this point, the corresponding ${\cal
PT}-$symmetric Hamiltonians cease to be tractable as equivalent to
any Hermitian ``physical" partner. In such a limiting case of the
theory, its genuine quantum meaning must be modified (or rejected).

Such a speculative idea moves us already beyond the scope of this
short review, with a useful intuitive guidance provided by the mere
toy example (\ref{twobytwoPT}), the singularity of the new metric of
which is easily verified by immediate calculation. Such an $a=b(t_0)$
model could still retain some information about the collapse of the
system at the critical time $t=t_0$.

In the same schematic example the third regime with $ |a| < |b|$
exhibits even more mind-boggling properties. Still the
time-development remains pseudounitary \cite{ich} and the smoothness
of transition offers a certain guidance in the not yet well-explored
domain where even causality may be put under question-mark. As we
mentioned, the explicit interest in this possibility characterizes
not only the older models in field theory \cite{Lee} but also some of
their innovated versions~\cite{KleefKy}.

In summary, simplified calculations open the path towards new
phenomenological models. In particular, one may work with models
using mere bosons (in place of more complicated fermions) in nuclear
physics \cite{Geyer} or speak about a generalized form of the ${\cal
CPT}-$symmetry within relativistic quantum field theory~\cite{BB2}
etc. In the future, perhaps, we shall be able to construct the models
where the pseudo-Hermiticity will prove necessary for a deeper or
more consistent treatment of the dynamical symmetries and/or laws of
evolution in quantized systems.

\section*{Acknowledgement}

Work supported by the GA AS grant Nr. A 104 8302.

\end{document}